\documentclass[]{aastex631}

\received{October 4, 2021}

\usepackage{multirow}  
\usepackage{amssymb}
\usepackage{gensymb}
\usepackage{amsmath}
\usepackage{xcolor}
\usepackage{ulem}

\submitjournal{Astrophys. J.}

\shorttitle{Linear Polarization}
\shortauthors{DeForest, Seaton \& West}
\graphicspath{{./}{figures/}}

\begin{document}

\title{Three-Polarizer Treatment of Linear Polarization in Coronagraphs and Heliospheric Imagers}

\author[0000-0002-7164-2786]{Craig E. DeForest}
\affiliation{Southwest Research Institute \\
1050 Walnut Street, Suite 300 \\
Boulder, CO 80302, USA}
\author[0000-0002-0494-2025]{Daniel B. Seaton}
\affiliation{Southwest Research Institute \\
1050 Walnut Street, Suite 300 \\
Boulder, CO 80302, USA}
\author[0000-0002-0631-2393]{Matthew J. West}
\affiliation{Southwest Research Institute \\
1050 Walnut Street, Suite 300 \\
Boulder, CO 80302, USA}

\correspondingauthor{Craig E. DeForest}
\email{deforest@boulder.swri.edu}

\begin{abstract}

Linearly polarized light has been used to view the solar corona for over 150 years.
While the familiar Stokes representation for polarimetry is complete, it is best matched
to a laboratory setting and therefore is not the most convenient representation 
either for coronal instrument design or for coronal data analysis. Over the last 100 
years of development of 
coronagraphs and heliospheric imagers, various representations have been used both
for direct measurement and analysis.  These systems include famous representations such
as the ($B$,~$pB$) system that is analogous to the Stokes system in solar observing
coordinates, and also internal representations such as in-instrument Stokes parameters
with fixed or variable ``vertical'' direction, and brightness values through a particular 
polarizing optic or set thereof.  Many polarimetric instruments currently use a symmetric 
three-polarizer measurement and representation system, which we refer to as ``($M,~Z,~P$)'', 
to derive the ($B$,~$pB$) or Stokes parameters.  We present a symmetric derivation of
($B,~pB$) and Stokes parameters from  ($M,~Z,~P$), analyze the noise properties of 
($M,~Z,~P$) in the context of instrument design, develop ($M,~Z,~P$) as a useful
intermediate system 
for data analysis including background subtraction, and draw a helpful analogy between 
linear polarimetric systems and the large existing body of work on photometric 
colorimetry.

\end{abstract}


\section*{~}
\begin{quote}
\textit{Light is a thing that cannot be reproduced but must be represented by something else: by color.}
~(Paul~C\'ezanne)
\end{quote}

\section{Introduction} \label{sec:intro}

The solar corona is linearly polarized \citep{arago_1843}. That property has been 
exploited over nearly a century of coronal observations: both for background removal
in coronagraphs \citep{lyot_1930} and for 3D analysis
\citep{poland_munro_1976,dekoning_pizzo_2011,deforest_etal_2017}.  

\begin{figure}
    \centering
    \includegraphics{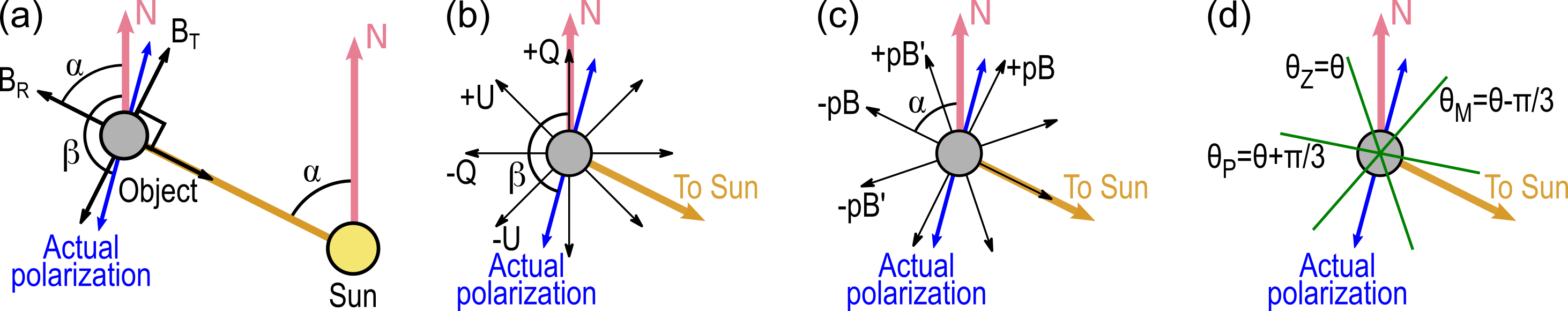}
    \caption{Four panels describe linear polarization analysis in a coronal context.  (a) An object near the Sun has position angle $\alpha$ and polarization angle $\beta$; $B_R$ and $B_T$ describe brightnesses through radially and tangentially aligned polarizers. (b) Stokes $Q$ and $U$ describe polarization in the ``+'' and ``$\times$'' directions relative to the instrument (or solar North).  (c) $pB$ and $pB'$ are Stokes parameter analogs in the solar observing reference frame.  (d) Observing polarization through three polarizers mutually separated by $\pi/3$ radians is sufficient to capture the polarization state ($I,Q,U$) or, equivalently, ($B,pB,pB'$). Although Thomson-scattered light is polarized in the $B_T$ direction, we show the polarization vector slightly misaligned, to emphasize the general case.}
    \label{fig:one}
\end{figure}

The degree of coronal polarization is conventionally reported via two parameters: a $B$
(``unpolarized  brightness'') parameter and its counterpart $pB$ (``polarized
brightness''); and both parameters may be mapped over an image plane to create separate
``$B$ images'' and ``$pB$ images''.  Because the corona is a distributed object, it is
best tracked via its radiance (delivered optical power per unit area, per unit solid angle),
and therefore ``brightness'' and ``radiance'' are synonyms.  The K-corona is visible 
primarily 
via Thomson scattering and therefore it is polarized perpendicular to the plane containing the observer, the
Sun, and the scattering point. At each point in a two dimensional 
image plane, the K corona is thus polarized perpendicular to a line extending
through the imaged center of the Sun and the given point
(Figure \ref{fig:one}a); this direction is not only predicted by theory
\citep[e.g.,][]{billings_1966,Howard_Tappin_2009}, but also readily verified through direct  
measurement \citep[e.g.,][]{Filippov_etal_1994}.  
$pB$ may thus be calculated as 
\begin{equation}\label{eq:def}
    pB = B_{T} - B_{R},
\end{equation}
where $B_{T}$ is the radiance observed through a linear polarizer oriented
tangentially to a solar-concentric circle passing through the image point of interest, 
and $B_{R}$ is the radiance observed through a linear polarizer
oriented radially to the Sun through the same point
\citep[e.g.,][]{Minnaert_1930,Altschuler_Perry_1972}. Whether
Equation \ref{eq:def} is accidental or fundamental to the definition of $pB$
is a matter of ongoing historical confusion and even dispute. This is explored 
in an Appendix to this article, which introduces a $^{\circ}pB$ and $^{\perp}pB$ 
to distinguish historical usage.  In this work we treat Equation \ref{eq:def} as fundamental;
this treatment implies that $pB$ is similar to the Stokes $Q$ parameter
\citep[e.g.,][]{hecht_zajac_1974}, but in a coordinate system that is rotated 
relative to the 
instrument.

The ($B,~pB$) representation of coronal polarization is convenient because it matches 
the observing geometry,
and in fact many coronagraphs have been constructed to measure directly that
element of linear polarization \citep[e.g.,][]{Altschuler_Perry_1972}. 
However, because the direction of ``radial'' varies in the image plane, this adds
complexity to the instrument itself. Recent spaceborne coronagraphs have instead measured
the full linear polarization state of incident light in the instrument frame
\citep[e.g.,][]{brueckner_etal_1995,howard_etal_2008}.  This requires capturing the first
three Stokes parameters ($I,~Q,~U$) at each location of the image plane, either directly or
indirectly via a representative measurement.

The Stokes ($I,~Q,~U$) parameters are defined in terms of exposures through individual
crossed
polarizers in a laboratory, and the textbook approach to measurement involves 
four exposures through crossed polarizers in the fixed instrument reference frame: horizontal/vertical 
for Q and diagonal for U
\citep[Figure \ref{fig:one}b;][]{hecht_zajac_1974}.  However, only three independent measurements
are necessary to determine the linear polarization states of a beam of
light.  
This was the basis of the triplet polarizing camera used by
\cite{ohman_1947} to observe
the 1945 eclipse, and \"Ohman attributed the technique to an earlier
analysis by
\cite{fesenkov_1935}.  The technique is described briefly in Chapter 4 of \cite{billings_1966}, and by \cite{newkirk_etal_1970}; these authors both
used the Stokes parameters as an intermediary system, rather than treating
the triplet analysis as primary.
Polarization triplet analysis has been and is still used routinely with coronal
images from Skylab \citep{poland_munro_1976}, SOHO/LASCO \citep{brueckner_etal_1995} and STEREO/SECCHI
\citep{howard_etal_2008}, and
is planned for other in-development instruments.

Despite its common use, full descriptions of the polarizer triplet 
technique, including 
instrumental effects, are elusive in the current literature; and those that exist
are somewhat unsatisfying because, in translating through the Stokes system,
the analytic work loses the symmetry of the polarizer triplet
system.  In this
article, we derive formulae to find directly
the ($B,~pB,~pB'$) or
Stokes ($I,~Q,~U$)
representations of linear polarization from polarizer triplet data; present an analytic noise analysis
for the most likely sources of noise or systematic error in a real instrument using a polarizer triplet;
introduce the use of an ($M,~Z,~P$) 
system of ``virtual polarizer triplets'' (Figure \ref{fig:one}d) to represent the linear polarization 
state of light
in frames other than the observing frame; and draw a surprising but helpful analogy 
between the major systems of linear polarimetry and corresponding systems for 
representing color.

\section{Definitions}\label{sec:defs}

Table \ref{tab:defs} defines several quantities, including radiances,
polarization parameters, and angles, that are relevant to a polarizing coronagraph or heliospheric imager.  These quantities are used throughout Section \ref{sec:calculations}, and are collected here for reference.  Figure \ref{fig:one} illustrates the angles and associated angles of the principal quantities in Table \ref{tab:defs}.

\begin{table}[tbh]
    \centering
    \begin{tabular}{c|c|c}
         \textbf{Quantity} & \textbf{Expression} & \textbf{Definition}  \\
         \hline
         \multirow{2}*{$B_{\{T,R\}}$} & \multirow{2}*{--} & Radiance through an ideal polarizer oriented \\ 
         ~ & ~ & tangentially or radially relative to the Sun \\
         \hline
         $B$ & $B_{T} + B_{R}$ & ``Unpolarized brightness'' (radiance) \\
         \hline
         $pB$ & $B_{T} - B_{R}$ & Coronal ``Polarized brightness'' \\
         \hline
         $\alpha$ & -- & Solar position angle of an image point \\
         \hline
         $\beta$ & -- & Direction of polarization  \\
         \hline
         $\theta$ & -- & Polarizer angle (also subscripted as $\theta_i$)\\
         \hline
         $\theta_i$ & $\theta+\{-1,0,1\}\pi/3$ & One of three angles in an ($M,~Z,~P$) triplet \\
         \hline
         $\phi$ & -- & Second polarizer angle \\
         \hline
         $B_{\theta}$ & -- & Radiance through a polarizer at angle $\theta$ \\
         \hline
         $B_{i}$ & -- & Radiance through a polarizer at angle $\theta_i$ \\
         \hline
         \multirow{2}*{$B_{\{|,-,/,\backslash\}}$} & \multirow{2}*{--} & Radiance through an ideal polarizer  \\
         ~ & ~ & oriented at $n\pi/4$ ($n\in\mathbb{Z})$ \\
         \hline
         \multirow{2}*{$I$} & $B_{|}+B_{-}$ & Stokes I -- synonym for $B$ (sum of \\
         ~ & $B_{\backslash} + B_{/}$ & radiance through any two perp. polarizers) \\
         \hline
         $Q$ & $B_{|} - B_{-}$ & Stokes Q (in instrument frame)\\
         \hline
         $U$ & $B_{\backslash} - B_{/}$ & Stokes U (in instrument frame) \\
         \hline       
         \multirow{2}*{$p$} & \multirow{2}*{$\sqrt{Q^2+U^2}/I$} & Polarization fraction (note: for \\
         ~ & ~ & Thomson scattering, $pB = (p)(B)$) \\
         \hline
         $S_i$ & $\sin\left[2\left(\theta_i-\alpha\right)\right]$ & Convenient abbreviation for sine expression\\
         \hline
         $C_i$ & $\cos\left[2\left(\theta_i-\alpha\right)\right]$ & Convenient abbreviation for cosine expression\\
    \end{tabular}
    \caption{Useful quantities in a linear polarimetric coronagraph.  In this context, Stokes $I$ is a radiance, not an intensity. Lists in curly braces enumerate options for the relevant expression.}
    \label{tab:defs}
\end{table}

\section{Linear Polarization from Three Polarizers}\label{sec:calculations}

We start by deriving a formula for $pB$ and $B$ at position angle $\alpha$, with coronal
light polarized
along the direction $\alpha+\pi/2$.  This further leads to expressions for extracting the
direction of
polarization, and the full linear portion of the Stokes representation.  Since $\alpha$ is
the angle of a radial line, the polarization direction $\beta$ is just $\alpha+\pi/2$
(tangential) for Thomson scattered light in the corona.

Solving the definitional expressions for $B$ and $pB$ in Table \ref{tab:defs}, for $B_T$ and $B_R$:
\begin{equation}\label{eq:btbr}
B_T=\frac{B+pB}{2} \hspace{0.3in}and\hspace{0.3in} B_R=\frac{B-pB}{2}
\end{equation}
Admitting the light through a polarizer at angle $\theta$ projects the electric field to the new direction; intensity is the
square of the electric field amplitude, so:
\begin{equation}\label{eq:btheta_brbt}
B_\theta = B_{T}\left[\sin\left(\theta-\alpha\right)\right]^2 + B_{R}\left[\cos\left(\theta-\alpha\right)\right]^2.
\end{equation}
Substituting and applying the double-angle cosine formula:
\begin{equation}\label{eq:btheta_bpb}
B_\theta = \frac{1}{2}\left\{
 B - pB \cos \left[2\left( \theta-\alpha \right)\right] 
\right\},
\end{equation}
which is the projection equation for partially polarized light.  Solving for $pB$ in terms of $B$ and $B_\theta$ yields:
\begin{equation}\label{eq:pb_onepol}
pB = \frac{B - 2B_\theta}{\cos\left[2\left(\theta-\alpha\right)\right]}~,
\end{equation}
which resolves $pB$ in terms of the radiance through a single arbitrary polarizer and also the 
total radiance with no polarizers in the beam.  Equation \ref{eq:pb_onepol} is problematic, because the denominator is small when $(\theta-\alpha)$ is near
$\pm\pi/4$.  If multiple polarizers are available at 
different angles, one can regularize 
Equation \ref{eq:pb_onepol} by taking an average, and weighting each term by the square of the 
offending cosine.  Adopting the $C_i$ abbreviation (Table \ref{tab:defs}):
\begin{equation}\label{eq:pb-npol}
pB = \frac{  \sum_i \left\{ \left( B - 2B_{i} \right) 
C_i \right\}}
{  \sum_i C_i^2}
\end{equation}
which is numerically stable if the $\theta_i$ are not all separated by intervals of $n\pi/2$.  
Choosing the $(M,Z,P)$\footnote{(for ``Minus, Zero, Plus'')} basis of polarizers 
at $\theta-\pi/3$, $\theta$, and $\theta+\pi/3$, the denominator
sum evaluates to 3/2 and the $B$ terms sum to zero, so:
\begin{equation}\label{eq:pb-3pol}
pB=-\frac{4}{3}\sum_{i\in\{M,Z,P\}}\left(B_{i} C_i\right)
\end{equation}
Meanwhile, solving Equation \ref{eq:btheta_bpb} for $B$ yields:
\begin{equation}\label{eq:b-1pol}
B = 2B_\theta + C\,pB, 
\end{equation}
where the $C$ is not subscripted because here it appears outside of a sum.  Equation \ref{eq:b-1pol} is not particularly useful by itself -- but averaging over the three polarizer positions
of the $(M,Z,P)$ system eliminates the cosine through a trigonometric identity, yielding:
\begin{equation}\label{eq:b-3pol}
B=\frac{2}{3}\sum_{i\in\{M,Z,P\}} B_{i}
\end{equation}
Equations \ref{eq:pb-3pol} and \ref{eq:b-3pol} give $B$ and $pB$ in closed and symmetrical
form, given radiance data through three polarizers with relative angles of $(-\pi/3)$, $(0)$, and $(+\pi/3)$
radians (60$\degree$ separation)
relative to a baseline angle $\theta$, with the assumption that the direction of linear polarization
is perpendicular to $\alpha$.  

Note that, although $\alpha$ is defined (in Table \ref{tab:defs}) as 
position angle around the Sun, nothing in Equations \ref{eq:pb-3pol} and \ref{eq:b-3pol}
requires any particular value of $\alpha$ -- or, for that matter, any particular orientation of the
main polarization direction $\theta$.  In fact, in the special case where $\alpha=\pi/2$, 
$B$ and $pB$ are just the Stokes $I$ and $Q$ parameters in the solar reference frame.  
Working by analogy to the Stokes parameters, we can substitute $\alpha\rightarrow\alpha+\pi/4$
into Equation \ref{eq:pb-3pol}; this yields a similar quantity that reduces to Stokes U in the
same circumstance:
\begin{equation}\label{eq:pbp-3pol}
pB'=-\frac{4}{3}\sum_{i\in\{M,Z,P\}}\left( B_{i} S_i\right)
\end{equation}
where the new quantity $pB'$ bears the same relationship to $pB$ as Stokes $U$ does to Stokes $Q$.
In systems where $pB'$ is important, we must generalize Equation \ref{eq:btheta_bpb} to include
$pB'$:
\begin{equation}\label{eq:btheta_bpbpbp}
B_\theta = \frac{1}{2}\left\{  B - C\,pB -
S\,pB'\right\}
\end{equation}
where the difference between Equation \ref{eq:btheta_bpb} and Equation \ref{eq:btheta_bpbpbp}
is that the former assumes tangential polarization while the latter treats arbitrary linear
polarization.  Repeating the derivation of Equation~\ref{eq:pb-3pol} using 
Equation~\ref{eq:btheta_bpbpbp} instead of \ref{eq:btheta_bpb} yields the same result, because the 
$C_i$ weighted summation in Equation~\ref{eq:pb-3pol} 
eliminates the $S_i\,pB'$ terms from Equation~\ref{eq:btheta_bpbpbp}.

The $B$,~$pB$,~$pB'$ system is an analog of the Stokes $I$,~$Q$,~$U$ system, but rotating with $\alpha$
around the Sun rather than being fixed in the instrument frame of reference; the familiar
Stokes parameters may be recovered anywhere by substituting $\alpha=\pi/2$ into Equations
\ref{eq:pb-3pol}, \ref{eq:b-3pol}, and \ref{eq:pbp-3pol}.  In the special case where the polarizer
triplet is aligned with the instrument, one may also set $\theta=0$ and arrive, after many 
cancellations, at the simplified
expressions:
\begin{equation}\label{eq:StokesQ2}
Q= \frac{2}{3}\left( 2B_Z-B_M-B_P \right)    
\end{equation}
and
\begin{equation}\label{eq:StokesU}
U=\frac{2}{\sqrt{3}} \left(B_P-B_M\right),
\end{equation}
which (together with the identity $B=I$) define the Stokes $I,Q,U$ triplet relative to the
reference Z polarizer in the $M,Z,P$ system, with a minimum of calculation. Equations \ref{eq:StokesQ2} and \ref{eq:StokesU} reproduce the derivations
presented by \cite{ohman_1947} and \cite{billings_1966}.

The remaining important system to represent linear polarization is $(B, \theta, p)$, which uses the 
overall brightness $B$, direction of polarization $\theta$, and normalized degree of polarization 
$p$.  The direction of polarization is available from $pB$ and $pB'$ (or, equivalently, from Stokes 
$Q$ and $U$).  Differentiating Equation \ref{eq:btheta_bpbpbp} to find the maximum value of $B_\theta$:
\begin{equation}\label{eq:derivative}
\left.\frac{\partial B_\theta}{\partial \theta}\right|_{\theta=\theta_{max}} = \left\{S\,pB - C\,pB'\right\} = 0~,
\end{equation}
so:
\begin{equation}\label{eq:angle}
\theta_{max} = \frac{1}{2}\arctan\left(\frac{pB'}{pB}\right)+\frac{\pi}{2}+\alpha~,
\end{equation}
where the four-quadrant arctan is implied and
selects maxima rather than minima, and the $\pi/2$ arises from canceling negative signs in the
numerator and denominator of the arctan (adding $\pi$ to the result of the four-quadrant arctan).  Making use of 
Equation \ref{eq:angle}, substituting the arctan into Equation \ref{eq:btheta_bpbpbp}, and 
recognizing two trigonometric cancellations yields the expected formula for $p$: 
\begin{equation}\label{eq:p}
p \equiv \frac{2B_{\theta,max}-B}{B} = \frac{\sqrt{pB^2 + pB'^2}}{B}.
\end{equation}
Equations \ref{eq:angle} and \ref{eq:p} recover the textbook formulas for $\theta_{max}$ and $p$ from 
the Stokes formalism, in the 
context of the $(B, pB, pB')$ system instead of $(I, Q, U)$.


\subsection{Error Sources in Three-Polarizer Polarimetry}

Polarimetry is affected by 
multiple error and noise sources that are specific to the 
measurement.  Here we derive expressions for calculating the error or noise associated
with three-polarizer measurements of $B$ and $pB$, given important photometric or mechanical
tolerances of the instrument.  The three major potential sources of error are
photometric noise, polarizer misalignment errors, and finite polarizer effectiveness. 
We consider noise as an approximately Gaussian-distributed random variable, with mean value of zero, added to the true value of each quantity.

To keep track of which quantities represent physical truth (with no noise nor measurement 
error) and which represent
inference from measurement (with noise and other error sources included), we introduce 
an overbar to indicate directly or indirectly measured parameters:  
\begin{equation}\label{eq:def-of-overbar}
\bar{X} \equiv X + \Delta X~,
\end{equation}
where $X$ is an observable parameter such as $B$ or $pB$, $\bar{X}$ is the observed value, and 
$\Delta X$ is a noise term.

Furthermore, we take the indices $i$ and $j$ to run over the
($M,~Z,~P$) polarizer angles when mentioned in a summation, in keeping with Equations \ref{eq:pb-3pol}, \ref{eq:b-3pol}, and \ref{eq:pbp-3pol}.

\subsubsection{Photometric errors}\label{sec:photometric-error}

Propagating photometric error and noise through multi-polarizer measurements is straightforward.
Photometry is most frequently limited by Poisson-distributed photon shot noise or similar uncorrelated noise sources.
Systematic errors can also contribute to total error; we neglect these terms, which is equivalent
to treating systematics in each channel as randomly distributed across channels. To each value
$B_i$ we add a noise term $\Delta B_i$ that we treat as a sample of a random variable. 
To propagate noise from the three samples, we take the partial derivative of Equations
\ref{eq:pb-3pol}, \ref{eq:b-3pol}, and \ref{eq:pbp-3pol} with respect to an arbitrary 
polarizer brightness, then sum the $\Delta B_i$ terms in quadrature.
This approach works because Equations \ref{eq:pb-3pol}, \ref{eq:b-3pol}, and \ref{eq:pbp-3pol} are linear in $B_i$; and the $B_i$ noise samples are also taken to be uncorrelated.
Equation \ref{eq:b-3pol} is straightforward to differentiate:
\begin{equation}\label{eq:partial-b-3pol}
\frac{\partial\, \bar{B}}{\partial \bar{B}_{i}} = \frac{2}{3}.
\end{equation}
In the most common case for image detectors, there is both a signal-independent 
noise component and a Poisson-statistics photon counting noise component, adding in quadrature.
In that case, we have
\begin{equation}\label{eq:Bi-composite-noise}
\Delta B_i = \sqrt{ (\Delta_{c}B)^2 + B_0 B_i },
\end{equation}
where $\Delta_{c}B$ is the signal-independent noise term (such as detector dark noise or read noise), $B_0$ is an instrument-specific 
constant of proportionality relating radiance units to detection quanta at an individual pixel or detector, and $\sqrt{B_0 B_i}$ is the Poisson noise term that arises from counting statistics. Merging
the noise
terms explicitly using Equation \ref{eq:b-3pol} and quadrature summmation yields
\begin{equation}\label{eq:B-composite-noise-sum}
\Delta B = \frac{2}{3} \sqrt{\sum_i\left[\left(\Delta_cB\right)^2+B_0B_i\right]}~,
\end{equation}
or, substituting from Equation \ref{eq:btheta_bpbpbp},
\begin{equation}\label{eq:B-composite-noise}
\Delta B = \frac{2}{3}\sqrt{ 3(\Delta_{c}B)^2 + \frac{B_0}{2} \sum_i \left( B - C_i pB - S_i pB'\right) }~.
\end{equation}
The $C_i$ terms and the $S_i$ 
terms sum to zero, leaving
\begin{equation}\label{eq:B-composite-noise-2}
\Delta B = \frac{2}{\sqrt{3}}\sqrt{(\Delta_c B)^2 + B_0 B/2}~, 
\end{equation}
which, in the special case of unpolarized light (so that $B_i=B/2$ for each $i$) would reduce, for each $i$, to
\begin{equation}\label{eq:B-noise}
\Delta B = \frac{2}{\sqrt{3}}\Delta B_i~,
\end{equation}
which arises from Equation \ref{eq:B-composite-noise-2} by substituting $B_i$ for $B/2$ and noticing that the sum matches the right-hand side of Equation \ref{eq:Bi-composite-noise}.

Because Equation \ref{eq:B-composite-noise-2} is independent of
$pB$ and $pB'$ for polarizer triplets, the photometric $\Delta B$ may be calculated for
the non-polarized case and remains valid in the polarized case, if the primary noise sources
scale independently of the signal and/or with Poisson counting statistics.  The form of Equation 
\ref{eq:B-noise} highlights that the photometric noise scales as expected from individual
samples being averaged together, shrinking by a factor of $3^{1/2}$ when the
polarizer triplet exposures are merged.

Turning to $pB$, the derivative of Equation \ref{eq:pb-3pol} is just:
\begin{equation}\label{eq:partial-pb-3pol}
\frac{\partial\, p\bar{B}}{\partial \bar{B}_{i}} = -\frac{4}{3} C_i.
\end{equation}
The quadrature sum is:
\begin{equation}\label{eq:pB-phot-noise-a}
\Delta pB = \frac{4}{3}\sqrt{ 
     \sum_{i}C_i^2 \Delta B_i^2
      }
= \frac{4}{3}\sqrt{\frac{3}{2}\left(\Delta_c B\right)^2 + B_0 \sum_i \left(C_i^2 B_i\right)}
\end{equation}
Substituting with Equation \ref{eq:btheta_bpbpbp} to eliminate the $B_i$ term yields
\begin{equation}\label{eq:pB-phot-noise}
\Delta pB = \frac{4}{3}\sqrt{\frac{3}{2}\left(\Delta_cB\right)^2 + \frac{3}{2}\frac{B_0B}{2} - \frac{B_0}{2}\left(pB\sum_iC_i^3 +pB'\sum_iC_i^2S_i\right)}
\end{equation}
where the summation terms oscillate in $\theta-\alpha$, with individual amplitude $3/4$ and
frequency 6 per full circle in $\theta$.  It is useful to bound the right-hand-side of Equation
\ref{eq:pB-phot-noise}.  The maximum value of the square root occurs when the oscillating term 
has its minimum (most negative) possible value.  That occurs at the extremal value of $pB$ 
and/or $pB'$, such that 
$pB^2+pB'^2 = B^2$.  Setting $pB = B$ and $pB' = 0$, and selecting the minimum value of the 
$C_i^3$ sum, yields
\begin{equation}\label{eq_pB-phot-noise-inequality}
\Delta pB \leq \frac{2\sqrt{2}}{\sqrt{3}}\sqrt{\left(\Delta_cB\right)^2+\frac{3}{4}B_0B}~,
\end{equation}
which can be simplified further.  Adding $(\Delta_cB)^2/2$ to the interior is allowed in the inequality, 
and allows a much simpler expression at the cost of slightly expanding the bound:
\begin{equation}\label{eq:pB-phot-noise-ineq-2}
\Delta pB \leq 2\sqrt{\left(\Delta_cB\right)^2 +B_0B/2} = \sqrt{3}\,\Delta B~.
\end{equation}
Systems that have nonzero values of $pB'$ have the same upper bound for $\Delta pB$, because 
when the total relative polarization
is maximized, mixing $pB$ and $pB'$ is equivalent to rotating the polarizers (changing $\theta$), 
which does not affect the minimization of the summation terms across $\theta-\alpha$. 

The upper bound for $pB'$ clearly must be the same as for $pB$, because the two quantities
are related by a rotation -- and neither $\theta$ nor $\alpha$ appear in Equation \ref{eq:pB-phot-noise-ineq-2}.  We therefore immediately write:
\begin{equation}\label{eq:pBp-phot-noise}
\Delta pB'  \leq \sqrt{3}\,\Delta B~.
\end{equation}

In sum: although the photometric noise level in the derived value of $pB$ 
(and, by extension, other related quantities $pB'$, $Q$, and $U$) from 
Equation \ref{eq:pb-3pol} varies with overall polarizer 
angle $\theta$, a simple upper bound exists that is independent of $\theta$ and (equivalently)
position angle $\alpha$.  Photometry in polarized brightness parameters
is worse than the unpolarized overall photometry, by a factor of up to $\sqrt{3}$.

\subsubsection{Polarizer Misalignments}

Instrument polarizers are mechanical devices and subject to alignment tolerance. 
The triplet $(M, Z, P)$
system (introduced between Equation \ref{eq:pb-npol} and Equations \ref{eq:pb-3pol} and
\ref{eq:b-3pol}) relies on a trigonometric identity to simplify the weighted average in
Equation \ref{eq:pb-npol}.  Errors in 
polarizer angle for a particular exposure translate directly to errors in $B$ and $pB$
independent of photometric noise.  Even assuming either perfect alignment or proper angular
calibration (using, e.g, Equation \ref{eq:pb-npol} or perturbation analysis on Equation
\ref{eq:pb-3pol}), individual polarized intensity images will be polarized at an angle that
differs slightly from nominal, due to noise in the alignment process.
Following the same methodology as in Section \ref{sec:photometric-error}, we characterize 
polarimetric response to this type of noise by treating noise-associated error 
in polarizer angle as a random variable.  We then propagate the noise from the foundational
equations to Equations \ref{eq:pb-3pol}, \ref{eq:b-3pol}, and \ref{eq:pbp-3pol}, by 
partially differentiating.

We start by characterizing $B_\theta$'s $\theta$-dependence.  From Equation \ref{eq:btheta_bpbpbp},
\begin{equation}\label{eq:dbthetadtheta}
\frac{\partial B_{i}}{\partial \theta_i} = S_i\,pB -  C_i\,pB'
\end{equation}
and, from Equation \ref{eq:b-3pol} (and making use of the same identity as in Equation \ref{eq:pb-3pol}),
\begin{equation}\label{eq:dbdtheta}
\frac{\partial \bar{B}}{\partial \theta_i} = \frac{2}{3}\left\{ S_i\,pB - C_i\,pB' \right\}
~.
\end{equation}
Treating three $\Delta\theta_i$'s in quadrature (and making use of the same identity as 
in Equation \ref{eq:pB-phot-noise}):
\begin{equation}\label{eq:delta-b-delta-theta}
\Delta B = \frac{\sqrt{2}}{\sqrt{3}}\sqrt{pB^2+pB'^2} \Delta \theta~.
\end{equation}
Equation \ref{eq:delta-b-delta-theta} describes a noise term in $B$ whose magnitude
is dependent on the 
total degree of linear polarization regardless of direction.  

Applying the same process to $pB$,  we  have from Equation \ref{eq:pb-npol} and Equation \ref{eq:dbthetadtheta}:
\begin{equation}\label{eq:dpbdtheta-A1}
\begin{split}
\frac{\partial p\bar{B}}{\partial\theta_i} = 
& \frac{1}{\sum_{k}C_{k}^2} \left\{ 
\sum_j
     (B-2B_{j})(-2S_i\delta_{ij})
     \right\}\\
     & + \frac{1}{\sum_{k}C_{k}^2} \left\{ \sum_j
     (2/3)(S_j\,pB - C_j\,pB')C_j
     \right\}\\
     & + \frac{1}{\sum_{k}C_{k}^2}\left\{ \sum_j
     \left(-2(S_j\,pB - C_j\,pB')\delta_{ij}C_j\right)
\right\} \\
&- \frac{2(C_i)(-2S_i)}{\left[\sum_{k}C_{k}^2\right]^2} \sum_j\left\{ (B - 2B_{j})C_j \right\}~,
\end{split}
\end{equation}
where $\delta_{ij}$ is the Kronecker $\delta$.  Substituting and, where possible, eliminating those terms that sum to zero:
\begin{equation}\label{eq:dpbdtheta-A2}
\frac{\partial p\bar{B}}{\partial\theta_i} = 
\frac{2}{3}\left\{ 
-2(B-2B_{i})S_i - 2(S_i\,pB-C_i\,pB')C_i +4C_iS_i\,pB
\right\}.
\end{equation}
Solving Equation \ref{eq:btheta_bpbpbp} for the quantity $B-2B_\theta$, and
substituting in to simplify Equation \ref{eq:dpbdtheta-A2} further:
\begin{equation}\label{eq:dpbdtheta-2}
\frac{\partial p\bar{B}}{\partial_{\theta_i}} = 
\frac{2}{3}\left\{
-2\left(C_i\,pB+S_i\,pB'\right)S_i - 2\left(S_i\,pB-C_i\,pB'\right)C_i + 4C_iS_i\,pB
\right\}~,
\end{equation}
or, gathering terms and applying the double-angle identity:
\begin{equation}\label{eq:dpbdtheta-4}
\frac{\partial p\bar{B}}{\partial{\theta_i}} = 
\frac{4}{3}\cos\left[4\left(\theta_i-\alpha\right)\right]pB'
~.
\end{equation}

Equation \ref{eq:dpbdtheta-4} describes a noise coupling term in $p\bar{B}$ that is
proportional
to $pB'$.  Therefore, because $pB'$ is negligible in the bright inner corona, small 
errors in polarizer
angle have negligible effect on the $pB$ calculations in that field.  This surprising 
result helps explain why triplet
polarization has performed well in historical coronagraph data: in
the inner few apparent solar radii of the corona, such systems are insensitive to
small polarizer misalignments.  
The corresponding error term is as curiously silent as a dog that does nothing in the night
\citep{doyle1892}. 

Applying three $\Delta\theta_i$'s in quadrature yields:
\begin{equation}\label{eq:delta-pb-delta-theta}
\Delta pB = \frac{2\sqrt{2}}{\sqrt{3}}\,pB'\,\Delta \theta~.
\end{equation}

Arguing from symmetry, we can also immediately write:
\begin{equation}\label{eq:delta-pbp-theta}
\Delta pB' = \frac{2\sqrt{2}}{\sqrt{3}}\,pB\,\Delta\theta~.
\end{equation}

In sum: polarizer misalignment in principle produces photometric error in derived
polarization parameters.  The resulting error in unpolarized brightness $B$
depends on total polarization and not, to first approximation order, on its direction.  The
corresponding error in $pB$ calculation depends only on the cross mode polarization $pB'$,
and vice versa.

\subsubsection{Polarizer Effectiveness}

The analysis leading to Equations \ref{eq:pb-3pol}, \ref{eq:b-3pol}, and \ref{eq:pbp-3pol} assumed 
perfect polarizers.  In practice, real polarizers are not perfectly effective and instead have a 
finite extinction coefficient.  A real instrument with physical polarizers will produce observed
polarizer brightnesses $\bar{B}_\theta$ with:
\begin{equation}\label{eq:btheta_bpbpbp_epsilon}
    \bar{B}_\theta = B_\theta + \epsilon B~,
\end{equation}
where $\epsilon$ is a small leakage coefficient (and of course an overall efficiency 
factor is 
removed via photometric calibration).  If $\epsilon$ is known, it can be calibrated out.
Equation \ref{eq:btheta_bpbpbp_epsilon} propagates into Equation \ref{eq:pb-3pol} as:
\begin{equation}\label{eq:pb-3pol-epsilon-i}
pB = -\frac{4}{3}\sum_i\left[\left(\bar{B}_{i}-\epsilon_i B\right)C_i\right]~.
\end{equation}
In the case where all the $\epsilon_i$ terms are equal, Equation \ref{eq:pb-3pol-epsilon-i} reduces to Equation
\ref{eq:pb-3pol}; i.e. finite polarizer extinction effects do not affect $pB$ 
measurement nor, by symmetry, $pB'$ measurement, provided that they are constant across polarizers.

Likewise, substituting $B=\bar{B}_i - \epsilon_iB$ into Equation \ref{eq:b-3pol}, we obtain:
\begin{equation}\label{eq:b-3pol-epsilon-i}
B  = \frac{2}{3}\frac{\sum_i\bar{B}_{i}}{\left(1+\frac{2}{3}\sum_i\epsilon_i\right)}~,
\end{equation}
so that, if all $\epsilon_i$ terms are equal, then the calculated $B_{i}$
value must be scaled by a factor of $(1-2\epsilon)$, to first approximation order, relative to the na\"ive perfect-polarizer formula in 
Equation \ref{eq:b-3pol}.

In sum: the polarizer extinction coefficient $\epsilon^{-1}$ does not affect inferred values of 
$pB$ nor of $pB'$, provided that it is constant across polarizer positions.  The extinction
coefficient does affect the inferred value of $B$, which can be corrected provided that 
$\epsilon$ is known.  As an example, polarizers with $\epsilon^{-1}=200$ would 
incur a $1\%$ error in the ratio $pB/B$ from this source alone, if not corrected using 
Equation \ref{eq:b-3pol-epsilon-i}.

\section{Virtual Polarizer Triplets}

So far we have established the theory of measuring the ($B$,~$pB$,~$pB'$) coronal polarization 
parameters or, equivalently, the ($I$,~$Q$,~$U$) Stokes parameters, using polarizer triplets.  But
the triplet formulation is useful not only as a means of measurement, but also as a means
to represent the linear polarization state of light.  A ``virtual polarizer
triplet'' formulation represents light as three brightness parameters through three ideal
polarizers separated by $\pi/3$ radians ($60^{\circ}$) each: the ($M$,~$Z$,~$P$) representation.  

In the context of coronal imaging, virtual polarizer triplets have
two distinct advantages over
the Stokes representation.  First, the virtual triplet is a symmetric representation:  the 
three brightness channels all have similar properties and none is treated preferentially, unlike
the Stokes and $pB$ systems in which Stokes $I$ (or $B$) is treated differently
from the other parameters.  Second, each parameter in the virtual triplet representation is
positive-definite; this is 
important in the context of coronal measurements, because it improves background subtraction when the background itself is polarized.  

Background subtraction for both coronagraphs and
heliospheric imagers requires accumulating and modeling a ``minimum background'' 
with varying degrees of sophistication depending on field of view
\citep[e.g.,][]{brueckner_etal_1995,howard_etal_2008}. For measurements close
to the Sun, the $pB'$ component is negligible and the background is nearly unpolarized
\citep[e.g.][]{wlerick_axtell_1957}.  At solar elongations of a few degrees, 
the polarization of the F corona becomes significant \citep{weinberg_hahn_1980} and its
polarization must be treated differently.  At solar elongations of 10$^{\circ}$ or more, 
the starfield itself is an important source of background light and must be eliminated
independently \citep{deforest_etal_2011}.  Adding insult to injury, the starfield is itself
slightly linearly polarized at up to the few-percent level by the interstellar medium
\citep[e.g.,][]{mignani_etal_2019}.  The positive definite nature of virtual polarizer images allows
the use of existing background estimation methods, which apply to unpolarized radiance, to be
used on each of the three polarizer channels independently.  This allows the use of existing 
techniques for removal of these background sources and instrument stray light
independently in each polarizer channel, to preserve the polarization as well as the 
overall brightness of the background signal \citep{deforest_etal_2017}.  Each virtual polarizer 
channel is itself ``just'' a radiance channel containing a fixed subset of the overall brightness
observed by the instrument; and existing techniques for finding a running-minimum brightness apply.

Because the starfield and the F corona are fixed in different coordinate systems, removing them
requires operating not only in different pixel coordinates \citep{deforest_etal_2011}, but also in
different polarization coordinates.  In particular, a linear polarization signal from a particular 
star will maintain a fixed direction in celestial coordinates but have variable direction of 
polarization in solar observing coordinates.  Therefore, coronal polarization values accumulated 
in an 
instrument coordinate system must be transformed into solar coordinates to identify and remove
the F corona; then into celestial coordinates to identify and remove the starfield; then back into
solar coordinates for further analysis.  This echoes the pixel resampling that is also necessary 
to eliminate those noise sources.  

In Section \ref{sec:calculations} we derived how to convert brightness values measured through
a triplet of polarizers, into either Stokes parameters (Equations \ref{eq:b-3pol},
\ref{eq:StokesQ2} and \ref{eq:StokesU}) or ($B$,~$pB$,~$pB'$) (Equations
\ref{eq:b-3pol}, \ref{eq:pb-3pol}, and \ref{eq:pbp-3pol}, respectively).  The inverse transform from either the 
Stokes or ($B,~pB,~pB'$) system to a virtual polarizer triplet at angle $\theta$ is given in 
Equation \ref{eq:btheta_bpbpbp}.  Here, we derive how to convert directly between $M$,~$Z$,~$P$ 
virtual triplet representations with different values of $\theta$.

Considering a polarizer at angle $\phi$, and substituting Equations \ref{eq:pb-3pol}, \ref{eq:b-3pol}, and \ref{eq:pbp-3pol} into Equation \ref{eq:btheta_bpbpbp}, gives:
\begin{equation}\label{eq:bphi-1}
B_\phi = \frac{1}{2}\left\{
\frac{2}{3}\left[\sum_iB_{i}\right] 
+ \cos\left[2\left(\phi-\alpha\right)\right]
\left[\frac{4}{3}\sum_i\left(B_{i}C_i\right)\right]
+ \sin\left[2\left(\phi-\alpha\right)\right]
\left[\frac{4}{3}\sum_i\left(B_{i}S_i\right)\right]
\right\}
~.
\end{equation}
Gathering terms across the finite sums gives
\begin{equation}\label{eq:bphi-2}
B_\phi = 
\frac{1}{3}\sum_i\left\{
B_{i} + 2B_{i}~\left(
    \cos\left[2\left(\phi-\alpha\right)\right]C_i
   +\sin\left[2\left(\phi-\alpha\right)\right]S_i
\right)
\right\}~,
\end{equation}
which simplifies to
\begin{equation}\label{eq:bphi-4}
B_\phi = \frac{1}{3}\sum_i
\left\{
B_{i}\left(1 + 2\cos\left[2\left(\phi-\theta_i\right)\right]\right)
\right\}
~,
\end{equation}
and thus to
\begin{equation}\label{eq:thetai-to-phi}
B_\phi = \frac{1}{3}\sum_{i\in\{M,Z,P\}}B_{i}\left[4\left(\cos\left[\phi-\theta_i\right]\right)^2-1\right]~.
\end{equation}
Equation \ref{eq:thetai-to-phi} yields the predicted brightness through a polarizer at angle $\phi$, given the brightness through
an ($M,~Z,~P$) triplet of three linear polarizers at angle $\theta$, i.e. a set of three linear polarizers at
$(\theta-\pi/3)$, $(\theta)$, and $(\theta+\pi/3)$.

Important properties of Equation \ref{eq:thetai-to-phi} include: (a) $B_\phi$ is by
construction
nonnegative for physical values of the $B_{i}$'s; (b) setting $\phi=\theta_i$ for 
any of the $i$'s recovers the relation $B_\phi=B_{i}$; and (c) the equation is numerically stable,
with no potential poles or singularities. 

Applying Equation \ref{eq:thetai-to-phi} to create predicted brightnesses for an ($M,~Z,~P$) triplet of 
polarizers at $(\phi-\pi/3$), $(\phi)$, and $(\phi+\pi/3)$ constitutes a change of basis, fully 
describing the polarization state not via direct brightnesses transmitted through a triplet of 
physical polarizers, but via predicted brightnesses through a triplet of ideal ``virtual polarizers'' at 
an arbitrary angle $\phi$.  Provided only that both $\theta$ and $\phi$ are known, the transformation 
via Equation \ref{eq:thetai-to-phi} preserves the full polarization state recorded by the original
triplet.

\section{Representations of Polarization and of Color}\label{sec:color}

There is a strong analogy between representations of color and representations of linear polarization. 

\begin{figure}[tbh]
    \centering
    \includegraphics{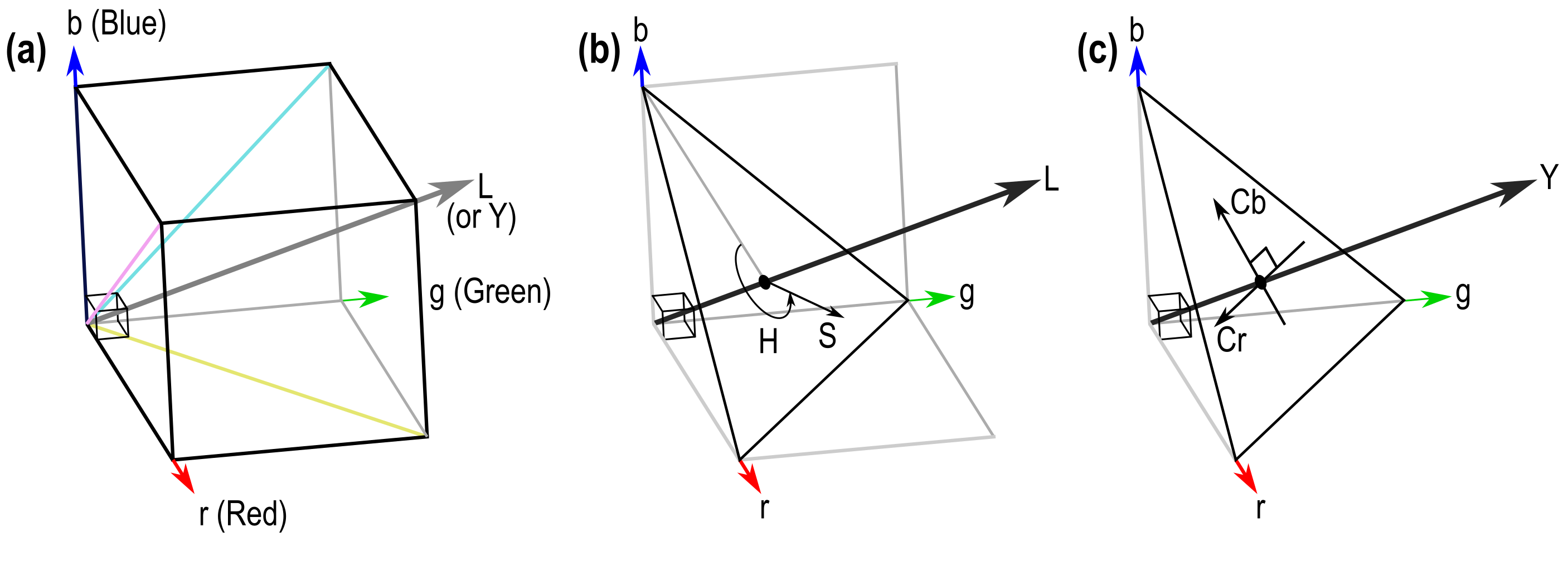}
\caption{Systems for representing color parameterize a 3-space that is analogous to the  3-space
needed to represent linear polarization.  (a) ($r,~g,~b$) uses conventional Cartesian coordinates 
to represent three separate radiances in different primary colors.  (b) ($H,~S,~L$) or ($H,~C,~L$) use 
conical or cylindrical coordinates, respectively, around a ``white line'' balancing $r$, $g$, and $b$, 
with $L$ and $C$ having units of radiance and $S\equiv C/L$.  
(c) ($Y,~Cb,~Cr$) uses Cartesian coordinates rotated to align with the white line.}
    \label{fig:two}
\end{figure}

The normal human visual system has three separate color channels at long, medium, and short wavelengths,
conventionally labeled ``red'', ``green'', and ``blue'', comprising a 3-D space  
\citep[e.g.,][]{young1802,maxwell1857,feynman1963}.  The components are conventionally labeled ($r,~g,~b$), 
with each representing radiance of light with a particular 
spectral characteristic (a
``primary color'' of the system, which forms a basis vector for the 3-space).
In practice, a large number of slightly different ($r,~g,~b$) systems exist, using slightly different 
primary colors.  All such spaces share the property of being positive-definite, 
like the radiance 
values discussed in Section \ref{sec:calculations}; this forms a positive-definite ``gamut''
that encompasses a subset of all 
possible colors (Figure \ref{fig:two}a). Representing colors outside the gamut would require
negative brightnesses of the primary colors, which is not physically possible.
Because of peculiarities of the human eye, no three
primary color vectors can both have positive-definite spectra (which may be reproduced
in the laboratory) 
and also encompass all humanly perceptible colors.  The
CIE 1931 ($X,~Y,~Z$) color space \citep{smith1931} is a standard ($r,~g,~b$) system that 
uses non-physical primary 
colors with negative
spectral intensities at some wavelengths, to provide a positive-definite gamut that 
encompasses all of human vision,
at the cost of requiring conversion to a physical ($r,~g,~b$) system before the color can be 
rendered for viewing.  
(Note that the $Z$ in ($X,~Y,~Z$) is distinct from the $Z$ in ($M,~Z,~P$): the former represents a blue-like fictional 
primary color, while the latter represents zero polarizer offset from a reference angle $\theta$).

The ($r,~g,~b$) systems are far from the only representations of color space.  Two other
large classes
are equally important.  

Hue systems are based on the \cite{munsell1912} color wheel, which parameterizes the space
as hue, ``chroma'', and ``luminance'' ($H,~C,~L$).  These systems are based around a
cylindrical projection
of ($r,~g,~b$) space. Luminance is a radiance of white light (distance from the 
origin along
a defined ``white line''
in colorimetric space, so that $r$, $g$, and $b$ have a given fixed ratio); 
chroma is a radial distance in the plane perpendicular to the white
line at particular radiance (the ``chroma plane''; 
and hue gives the angle at which the radius is to be drawn in that plane 
(Figure \ref{fig:two}b).
A more common variant of the ($H,~C,~L$) system is ($H,~S,~L$); this system replaces 
$C$ (chroma) with a 
relative value $S$ (``saturation'') defined via $S\equiv C/L$, and L (luminance) 
with a closely related quantity ``lightness''.

``Opponent'' color systems work similarly to ($H,~C,~L$) systems: they represent color
with luminance (usually $Y$ from the ($X,Y,Z$) system) and 2-D signed chroma values in the
chroma plane.
The ($Y,~Cb,~Cr$) system \citep{itu2017} has an unsigned
luminance signal indicating distance from the origin along the white line, and two signed
``chrominance'' signals (with units of radiance) that describe location in the perpendicular
plane (Figure \ref{fig:two}c). Here, $Y$ represents luminance and is normally a linear
combination of the 
($r,~g,~b$) signals: $Y\equiv K_r\,r+K_g\,g+K_b\,b$, with $\sum_iK_i^2=1$. For convenience in this
cursory treatment, we ignore this scaling, effectively setting $Y=L$ and $K_i=1/\sqrt{3}$ for all
the $K$s.  The ($Y,Cb,Cr$) and related color-opponent systems are important both for video coding
and because they describe well the perceptual aspects of the human visual system, as can be
verified by direct visual experiment \citep{churchland2005}.

Converting from $(r,~g,~b)$ to $(Y,~Cb,~Cr)$ is a simple linear transformation.  
With 
$\hat Y \cdot \hat r =\hat{Y}\cdot\hat{g}=\hat{Y}\cdot\hat{b}$ by construction, the realization that
the chroma-plane-projected ($\hat{r}',~\hat{g}',~\hat{b}'$) vectors are separated by $\pi/3$, and direct evaluation of 
$\cos(\pi/3)$ and $\sin(\pi/3)$, the coefficients may be 
written down by simple inspection:
\begin{equation}\label{eq:Y-from-rgb}
    Y=\frac{2}{3}\left[r+g+b\right]~,
\end{equation}
\begin{equation}\label{eq:Cb-from-rgb}
    Cb=\frac{1}{\sqrt{3}}\left[2b - r - g\right]~, \textrm{and}
\end{equation}
\begin{equation}\label{eq:Cr-from-rgb}
    Cr=\frac{2}{\sqrt3}\left(r - g\right).
\end{equation}
The $Cb$ and $Cr$ formulae describe perpendicular vectors in the chroma plane.  
The $Cb$ formula arises because of the symmetry of the triangle formed by the projected
($\hat{r}',~\hat{g}',~\hat{b}'$) directions in the chroma plane, and the fact that $\hat{Cb}$ is parallel to $\hat{b}'$. 
The $Cr$ formula is simpler because $\hat{Cr}$ is parallel to the $\hat{g}'-\hat{r}'$ line and perpendicular to $\hat{b}'$.

Equations \ref{eq:Y-from-rgb}, \ref{eq:Cb-from-rgb}, and \ref{eq:Cr-from-rgb} echo Equations \ref{eq:b-3pol},
\ref{eq:StokesQ2}, and \ref{eq:StokesU} respectively, establishing the analogy between ($r,~g,~b$) colors and 
polarizer triplet brightnesses --- and, similarly, between ($Y,Cb,Cr$) colors and the linear Stokes parameters.

Moving to the $(H,~S,~L)$ system, measuring the hue angle relative to $\hat{b}'$ immediately gives: 
\begin{equation}\label{eq:hue}
    H = \arctan\left[Cr/Cb\right]~,
\end{equation}
where the four-quadrant arctan is implied, and
\begin{equation}\label{eq:saturation}
    S = \frac{\sqrt{Cr^2+Cb^2}}{Y}~.
\end{equation}
Again, this echoes directly the structure of Equations \ref{eq:angle} and 
\ref{eq:p}, converting from the Stokes formulation to the ($B,\theta,p$) system, with $Y$, $H$, and $S$ corresponding to 
$B$, $2\theta'$, and $p$ respectively (where $\theta'$ has the appropriate reference angle subtracted 
for either the $(I,~Q,~U)$ or 
$(B,~pB,~pB')$ system).

Two useful insights come immediately from the connection between polarization and color spaces.  The first is an
understanding of why polarizer triplets are convenient and numerically stable representations of polarization space:
they are simply 3-vectors in the space, forming an orthonormal basis.  Polarizer triplet (or virtual
polarizer triplet) brightness parameters have a higher degree of mutual symmetry than do
Stokes parameters
(which are themselves a different orthonormal basis of the same space), but as orthornormal
bases they are easily analyzed, 
manipulated, and interconverted.

The second useful insight is the
notion of a ``physical gamut'' -- the range of values that are both representable, and also not physically
impossible, within a given 
representation space.  Because radiances of light cannot be negative, and the spectra of the
human eye's photosensitive
pigments (treated as very-high-dimensional vectors) are not perpendicular, no physically
realizable primary colors can
represent all of human color in a tricolor system.  The specific ($X,~Y,~Z$) system of
($r,~g,~b$) primary colors
\citep{smith1931} uses impossible (non-physical) spectra to achieve a gamut that can
represent any possible color of 
light, at a cost of requiring conversion to a specific physical colorimetric system before
color can be rendered in the physical world.  

A peculiarity of ($X,~Y,~Z$) is that it can also represent physically impossible colors 
that cannot exist at all, even in principle, because they cannot be rendered with any
positive-definite spectrum of light.  This is also the 
case for ($M,~Z,~P$) systems for representing polarization:
some of the possible vectors yield values of $pB$ and $pB'$ (or, equivalently, Stokes 
$Q$ and $U$) that are outside the Stokes
inequality $B^2 \ge pB^2+pB'^2$ which determines physically possible polarization states.  
In the ($X,~Y,~Z$) system, the physically realizable colors
occupy an irregularly shaped locus in the chroma plane, driven by the specific absorption
spectra in the human eye.  In an ($M,~Z,~P$) system, the physically realizable polarizations (that satisfy the Stokes inequality)
form a circle in the ``polarization plane'' described by ($pB,~pB'$) or ($Q,~U$); 
the circle is a cross-section of the Poincar\'e sphere \citep{bornwolf1999}, neglecting the 
third Stokes perturbation parameter $V$.

Happily, the circle is inscribed in the triangle formed by the ($M,~Z,~P$) gamut on the
polarization plane in polarimetric space, as can be verified by direct substitution into Equation \ref{eq:btheta_bpbpbp}.  This triangle is exactly analogous to the 
triangle in the chroma plane formed by the ($r,~g,~b$) gamut, which may be seen in Figure 
\ref{fig:two}, panels (b) and (c).

The presence of the 
``Poincar\'e circle'' physical gamut inside the triangular ($M,Z,P$) gamut ensures that all valid polarization
states may be represented by any suitable ($M,Z,P$) triplet of polarizers (as in Equation
\ref{eq:thetai-to-phi}), because the circle maps to itself under rotations about the center point. 

Many
impossible polarizations (which do not satisfy the Stokes inequality and 
 are therefore outside the Poincar\'e circle) may be represented with a particular
 polarizer triplet.
Substituting such triplet values into Equation \ref{eq:thetai-to-phi} may yield negative polarizer radiances in the new system.  As a trivial example, a point 
with $B_{\theta,M}=B_{\theta,P}=0$ and $B_{\theta,Z}=1$  yields negative values
of $B_\phi$ for nearly all values of $\phi-\theta$, because no physically realizable
polarization state can produce that combination of radiances through an ideal linear
polarizer triplet.  Contrariwise, all physically realizable states exist within the Poincar\'e circle, which is inscribed in all possible polarizer-triplet gamuts on the polarization plane, and therefore those states all yield positive values from Equation \ref{eq:thetai-to-phi} regardless of the value of $\phi-\theta$.

In sum: the principal systems of representing linear polarization are mathematically
analogous to
the principal systems of representing color.  This yields both intuition, via the 
connection to ($r,~g,~b$) color systems, and insights into why the polarizer triplet
system works well for representing linear polarization of light.

\section{Conclusions}

Linear polarization is an important measurement of the solar corona and
heliosphere.  In the solar case, the Stokes parameters (I,Q,U), which are
defined in the instrument frame of reference, are not the most convenient
measure of polarization; instead, because of the peculiarities of polarization
in the visible solar corona, a large body of past and current literature 
treats the polarized
brightness as a sort of ``virtual Stokes parameter'' oriented relative to
the image-plane solar radial direction rather than in a particular 
instrument direction, with its Stokes complement $pB'$ generally ignored.  
Building on that practical and implicit treatment,
we have presented the
derivation of the Stokes-equivalent $B$, $pB$, and $pB'$ parameters 
directly from a triplet of brightness values in three polarizers.  While
the three-polarizer approach to measuring coronal polarization was described
(in Russian) by \cite{fesenkov_1935} and more briefly (in English) by
\cite{ohman_1947} and \cite{billings_1966}, and is now commonly used in many
instruments, the derivation has generally used an asymmetric transition
through the Stokes parameters and/or been given cursory treatment. 

Instruments making use of a polarizer triplet are subject to particular
uncertainties in measurement that are peculiar to the polarizer triplet system.  
A literature search found no complete analytic 
treatment of these in-principle knowable noise sources.  Accordingly, 
we have presented the effects of three major
error or noise sources in three-polarizer instruments, on polarimetry derived
using the formulae presented here.  These error sources are: per-channel
photometric error; polarizer orientation error; and finite polarizer
extinction
coefficient.  The most surprising result is that polarizer orientation error 
does not affect the $pB$ measurement in first approximation order, in the inner 
few solar radii of the solar corona where total polarization is in the tangential 
($pB$) direction.  Our noise results supply indicative limits relating these major noise sources in polarization measurements, to noise in the calculated quantities $B$, $pB$, and $pB’$; but we did not consider non-Gaussian probability distributions of the input data, nor noise sources with significant covariance across measurements.  Further, specific definitions of derived parameters can affect noise terms in subtle ways for derived quantities that are not linearly related to the source quantities.  For example, even ``well-behaved’’ zero-mean, Gaussian-distributed photometric noise (such as we considered here) produces a net offset (non-zero mean perturbation) in determination of the total polarized brightness, $^\circ pB \equiv (Q^2+U^2)^{0.5} = (p)(B)$, due to geometrical effects in parameter space \citep{Inhester_etal_2021}

The three-polarizer representation of polarization is symmetric and readily
transformed to different orientations.  Using 
``virtual polarizer triplets'', to represent polarization data symmetrically 
in multiple frames, conveniently maintains the positive-definite and
symmetric properties of polarizer triplet 
measurements without 
corresponding
physical polarizers.  The mathematical 
transformation 
between different three-polarizer systems is readily evaluated and 
numerically stable, making virtual polarizer triplet analysis a useful
approach for converting polarization data between different coordinate
systems.  

Possible applications 
for a virtual polarizer triplet representation include
regularizing
polarimetry from a coronagraph or heliospheric imager that orbits the Sun or 
Earth, and subtracting multiple types of data background that are 
polarized in
different coordinate systems relative to the instrument.  Three such overlapping backgrounds are
potentially-polarized instrument stray
light (fixed in the instrument frame), wide field F corona (fixed in the solar
frame), and the starfield (fixed in the celestial frame).  

There is a direct mathematical analogy between representations of linear
polarization, and representations of visual color; this analogy 
may help guide intuition for the analytic derivations presented
here and potentially other applications of polarization space. 
The ($M,~Z,~P$), ($B,~pB,~pB'$) and Stokes, and ($B,~\theta,~p$) 
systems of representing linear polarization are seen to be direct analogs of the 
($r$,~$g$,~$b$), ($Y$,~$Cb$,~$Cr$), and ($H$,~$S$,~$L$) systems for representing 
color.  With
this understanding, it becomes clear why the triplet polarizer system works well for 
representing and manipulating polarization values.  The triplet brightness values are 
immediately seen to form
an orthonormal basis of the polarization space; and physically allowable polarization 
states are seen to be representable independent of reference angle, 
because of the circular
geometry of the Poincar\'e sphere.

\begin{acknowledgements}
The authors thank A. Caspi, M. Beasley, and C. Lowder for useful 
discussion and 
review of the article.  The analysis and discussion were improved by
insights from the anonymous referee.  This work was funded through PUNCH, 
a NASA Small Explorer
mission, via NASA Contract No. 80GSFC18C0014.

\end{acknowledgements}

\bibliography{polarization}{}
\bibliographystyle{aasjournal}

\section{Appendix: definitions of pB}\label{sec:Appendix}

Historically, the fact that the corona is tangentially polarized has
led to ambiguity in the term $pB$, which depending on context is used to refer 
either to total polarization or to only a certain component of polarization.
In the former case, authors define and use a quantity like $^{\circ}pB \equiv (Q^2+U^2)^{1/2}$ (where $Q$ and $U$ are the relevant Stokes parameters and 
Stokes $V$ is neglected). $^{\circ}pB$ is thus the total polarized brightness 
regardless of direction, and is  
analogous to $C$ in the ($H,C,L$) color system in Section \ref{sec:color}.  In the latter case, authors define and use a quantity like
$^{\perp}pB \equiv B_{T}-B_{R}$, i.e. the difference between the radiance observed 
through a tangentially-aligned and a radially-aligned linear polarizer.  $^{\perp}pB$ is effectively a variant on the Stokes parameter $Q$, oriented radially in a solar image plane (Figures \ref{fig:one}a and \ref{fig:one}c).

The distinction between $^{\circ}pB$ and $^{\perp}pB$ is nearly moot in lower coronal 
studies, where 
observed light happens to be polarized perpendicular to the focal-plane solar radial direction, though $^\circ pB$ responds counter-intuitively to photometric noise \citep{Inhester_etal_2021}; but
the two quantities generalize differently in
cases where the overall polarization doesn't happen to be tangential -- for example, in 
wide field coronagraphs or polarizing heliospheric imagers, where non-coronal polarized
light sources are significant sources of background radiance. In addition to instrumental stray
light, two such sources are the zodiacal light \citep{leinert_etal_1981},
and the starfield itself \citep{heiles_2000}.

Use of the $^{\circ}pB$ formalism is widespread in the coronal
literature.  Examples include
\citet{ohman_1947} and \citet{billings_1966}, who define an $I_p$ similarly to our
$^{\circ}pB$. \citet{MunroJackson1977} used $^{\circ}pB$ in their analysis of the 
corona, apparently to avoid having to deal with angle explicitly (based on their
Equation 3, which treats Thomson scattering).  More recently, the SOHO/UVCS
\citep{Kohl_etal_1995}, SOHO/LASCO \citep{brueckner_etal_1995}, and STEREO/SECCHI \citep{howard_etal_2008} analysis pipelines
calculated $^{\circ}pB$ explicitly \citep[e.g.,][]{Ofman_etal_2000,Dere_etal_2005}; 
and sources as recent as \citet{Vorobiev_etal_2014} and \citet{Reginald_etal_2017} 
use the formalism in describing new polarimetric techniques using multiplexed
single-frame polarimetric detectors.

The $^{\perp}pB$ formalism is equally widespread through the coronal literature, and 
seems to extend even earlier than $^{\circ}pB$.  \citet{Minnaert_1930}, in his
seminal work on coronal polarimetry, follows \citet{young_1911} in defining
the quantity $p\equiv(J_t-J_r)/(J_t+J_r+2A)$, in which $p$ is relative polarization,
$J_t$ and $J_r$ are radially and tangentially polarized elements of the visual corona,
and $A$ is an unpolarized background brightness.  We can immediately recognize 
that the quantity $(J_t+J_r+2A)$ is more commonly abbreviated $B$, and that 
Minnaert's $p(J_t+J_r+2A)$ is an expression for $^{\perp}pB$.  The HAO K coronameter
\citep{Altschuler_Perry_1972} was specifically built to exploit the tangential direction
of coronal polarization, and $^{\perp}pB$ appears explicitly (as $pB$) in their analysis. 
Additional important references using $^{\perp}pB$ span six decades
and include \citet{Saito_1965}, \citet{koomen_etal_1975}, \citet{crifo_etal_1983},
\citet{Hayes_etal_2001},
\citet{MoranDavila2004}, \citet{Howard_Tappin_2009}, \citet{dekoning_pizzo_2011},
\citet{deforest_etal_2013}, \citet{Dai_etal_2014}, and \citet{deforest_etal_2017}.  
All of these sources define a $^{\perp}pB$ and refer to it as ``polarized brightness'';
most refer to the quantity as $pB$ although some use $I_p$ instead (note that $I_p$ is used in these sources as a synonym for $^{\perp}pB$ and in other sources, mentioned above, as a synonym for $^{\circ}pB$).

The distinction between $^{\circ}pB$ and $^{\perp}pB$ has been so blurred, 
that some authors might even use both in the same paper, without acknowledgement of
the distinction.
For example, both
\citet{guhathakurta_etal_1999} and \cite{Frazin_etal_2012} compare $pB$ values
across different instruments, some of which generate $^{\circ}pB$ data products and some
of which generate $^{\perp}pB$ data products.  

The polarized brightness parameter $pB$ is well entrenched in the literature, 
and the two meanings $^{\circ}pB$ and $^{\perp}pB$ are close enough to not differ greatly
in the inner corona.  However, in wide-field coronagraphs and polarizing heliospheric imagers,
other polarized light sources come into play.  In these contexts the two parameters differ
greatly in meaning; past muddiness requires more clarity from present and future authors, who
ought to define $pB$ explicitly (as many do already) wherever it is used.

\end{document}